\documentclass[11pt,letterpaper]{article}

\usepackage{amssymb}
\usepackage{amsmath}

\usepackage{graphicx}

\usepackage[round,numbers,sort&compress]{natbib}

\newcommand{\Dl}{{D^{\rm c}_{\rm L}}}
\newcommand{\Dlt}{{D^{\rm c}_{\rm L,T}}}
\newcommand{\Dt}{{D^{\rm c}_{\rm T}}}
\newcommand{\eg}{{e.g., }}
\newcommand{\etam}{{\eta_{\rm m}}}
\newcommand{\etaf}{{\eta_{\rm f}}}
\newcommand{\etaeff}{{\eta_{\rm m}^{\rm eff}}}
\newcommand{\etal}{{\it et al.} }
\newcommand{\ie}{{i.e., }}
\newcommand{\kT}{{k_{\rm B}T}}
\newcommand{\pd}{{\partial}}
\newcommand{\rhat}{\hat{r}}
\newcommand{\tenG}{{\bf G}}
\newcommand{\tG}{\tilde{G}}
\newcommand{\vecf}{{\bf f}}
\newcommand{\vecF}{{\bf F}}
\newcommand{\vecL}{{\bf L}}
\newcommand{\vecOmega}{{\bf\Omega}}
\newcommand{\vecq}{{\bf q}}
\newcommand{\vecr}{{\bf r}}
\newcommand{\tenS}{{\bf S}}

\newcommand{\vecU}{{\bf U}}
\newcommand{\vecv}{{\bf v}}
\newcommand{\xhat}{\hat{x}}

\begin{document}

\title{Correlated Diffusion of Membrane Proteins\\ and Their Effect
on Membrane Viscosity}

\author{Naomi Oppenheimer and Haim Diamant\thanks{
\scriptsize Corresponding author. Tel.: +972 3 640 6967, Fax: +972 3 640 9293,
E-mail: hdiamant@tau.ac.il} \\~\\
School of Chemistry, Raymond \& Beverly Sackler Faculty of Exact Sciences,\\
Tel Aviv University, Tel Aviv 69978, Israel}

\date{September 24, 2008}

\pagestyle{myheadings}
\markright{Correlated Diffusion of Membrane Proteins}

\maketitle

\abstract{
  
  We extend the Saffman theory of membrane hydrodynamics to account
  for the correlated motion of membrane proteins, along with the
  effect of protein concentration on that correlation and on the
  response of the membrane to stresses. Expressions for the coupling
  diffusion coefficients of protein pairs and their concentration
  dependence are derived in the limit of small protein size relative
  to the inter-protein separation. The additional role of membrane
  viscosity as determining the characteristic length scale for
  membrane response leads to unusual concentration effects at large
  separation---the transverse coupling increases with protein
  concentration, whereas the longitudinal one becomes
  concentration-independent.

~
}


\section*{Introduction}

Biomembranes contain a high concentration of proteins, performing key
cellular functions \cite{biology}. Extensive efforts have been
directed, therefore, at measuring the dynamics of membrane proteins
using various experimental techniques \cite{Lippincott-Schwartz2001}.
Those studies have concentrated on either single-molecule dynamics
(using single-molecule tracking \cite{McConnell,Kusumi} and
fluorescence correlation spectroscopy \cite{Lippincott-Schwartz2001})
or collective gradient diffusion (using fluorescence recovery after
photobleaching \cite{FRAP}). Both the single-particle and large-scale
levels are in some contrast with the expected cooperative motion of
several proteins at the high concentration relevant to biomembranes,
and the crucial role played by small protein aggregates in membranes.
Until now experimental investigations of the correlated motion in
membranes and monolayers have been limited to rather extended objects,
such as domains \cite{Klinger1993,Cicuta2007} and embedded
microspheres \cite{Prasad2006}, although similar two-point
microrheological measurements with fluorescent membrane proteins seem
feasible \cite{McConnell}.  Furthermore, in cases where one is
interested in membrane properties rather than those of the proteins,
two-point microrheology has the advantage of being insensitive to the
shape of the inclusion and the local perturbation that it introduces
in the membrane \cite{Crocker2000,Naji2007}. It seems to be of
significant interest, therefore, to account for the correlated
Brownian motion of membrane proteins and the associated effects on
membrane dynamics.

From a hydrodynamic perspective, a bare, protein-free membrane can be
viewed as a quasi-two-dimensional (quasi-2D) liquid, whose molecules
(lipids) are free to flow only within the membrane surface, yet
exchange momentum with the surrounding three-dimensional (3D) liquid.
Hence, flows within a membrane are essentially different from those in
both 3D and 2D liquids, in that they do not conserve momentum. As a
result, a characteristic length scale $\kappa^{-1}$ emerges, such that
over distances much smaller than $\kappa^{-1}$ the membrane keeps its
momentum and responds similar to a 2D liquid, whereas beyond that
distance momentum is exchanged with the surroundings, and the response
is significantly modified.

To gain further intuition for the results that will follow, let us
compare this situation in slightly more detail with the ones in
momentum-conserving 3D and 2D liquids.  In the 3D case the stress (\ie
momentum flux) emanating from a local perturbation must decay with
distance $r$ as $\sigma\sim r^{-2}$ (so that the total flux through an
envelope of radius $r$ should remain constant). Since shear stress is
related to fluid velocity $v$ as $\sigma\sim\etaf\nabla v$, where
$\etaf$ is the shear viscosity of the fluid, the velocity response to
the perturbation decays as $v\sim(\etaf r)^{-1}$. More specifically,
given a point force $\vecf$ acting on the liquid at the origin, the
steady-state flow velocity $\vecv(\vecr)$ at position $\vecr$ is given
by \cite{Happel}
\begin{equation}
  \mbox{3D liquid:}\ \ v_i(\vecr) = G_{ij}(\vecr)f_j,\ \ 
  G_{ij}(\vecr) = \frac{1}{8\pi\etaf r} \left( \delta_{ij} 
  + \frac{r_ir_j}{r^2} \right),
\label{Oseen3D}
\end{equation}
where $\tenG$ is the Oseen tensor and $i,j=x,y,z$. (Summation over
repeated indices is implied throughout this article.) An important
consequence of this velocity response is that the correlation
(hydrodynamic interaction) between the motions of two particles
suspended in the liquid is long-ranged, decaying as $1/r$
\cite{Happel}.  Similarly, in a 2D momentum-conserving liquid the
stress must decay as $\sigma\sim r^{-1}$, and, thus, the velocity
decays as $-\eta_{\rm m}^{-1}\ln(\kappa' r)$, where $\etam$ is the 2D shear
viscosity. The resulting well known logarithmic divergence of this
problem \cite{Happel} requires a cutoff length $\kappa'^{-1}$ (\eg the
system size).  The analogue of Eq.\ \ref{Oseen3D} for the 2D case is
\begin{equation}
  \mbox{2D liquid:}\ \ v_i(\vecr) = G_{ij}(\vecr)f_j,\ \ 
  G_{ij}(\vecr) = \frac{1}{4\pi\etam} \left[ -\ln(\kappa' r) \delta_{ij} 
  + \frac{r_ir_j}{r^2} \right],
\label{Oseen2D}
\end{equation}
where here (and in the rest of the article) $i,j=x,y$. For the
intermediate quasi-2D case of fluid membranes we expect the response, 
and thus the hydrodynamic interaction between inclusions, to cross
over from the 2D logarithmic decay at short distances to the 3D
$1/r$ decay at large distances.

Three major theoretical approaches to membrane hydrodynamics have been
presented. In Saffman's pioneering theory \cite{Saffman} the membrane
is modeled as a flat slab of viscous liquid, having width $w$ and
viscosity $\etam/w$. The flow velocity in the membrane is assumed to
be two-dimensional, \ie the velocity profile across the slab width is
uniform. The slab has at its two bounding surfaces no-slip contacts
with an infinite viscous fluid (water), having viscosity $\etaf$.  The
emergent characteristic length---the Saffman-Delbr\"uck length
\cite{Saffman&Delbruck}---is given by
\begin{equation}
  \kappa^{-1} = \frac{\etam}{2\etaf}. 
\label{kappa}
\end{equation}
(In asymmetric cases, where the liquids on the two sides of the
membrane have different viscosities, one should replace $2\etaf$ with
the sum of those viscosities.)  Considering a membrane protein as a
cylindrical inclusion of radius $a\ll\kappa^{-1}$, the Saffman theory
yields the following self-mobility for the protein,
\begin{equation}
  B_{\rm s} = \frac{1}{4\pi\etam}\left[-\ln(\kappa a/2)-\gamma\right],
\label{SD}
\end{equation}
where $\gamma\simeq 0.58$ is Euler's constant.  The self-diffusion
coefficient of the protein is simply given, through Einstein's
relation, by $D_{\rm s}=\kT B_{\rm s}$, where $\kT$ is the thermal
energy. The weak logarithmic dependence of $D_{\rm s}$ on the protein
size, as predicted by Eq.\ \ref{SD}, does not seem to be obeyed in
practice \cite{Gambin2006}, possibly because of the local disturbance
that the protein creates in the membrane \cite{Naji2007}. Equation
\ref{SD} has been confirmed, nonetheless, in the Brownian motion of
small membrane domains \cite{Cicuta2007}. The Saffman theory has been
extended to arbitrary values of $\kappa a$ \cite{Hughes1980} and to
the case of membranes supported on a liquid layer of finite thickness
\cite{Stone1998}.

In the second approach, introduced by Levine \etal
\cite{Levine2002,Levine2004a,Levine2004b}, the membrane is treated as
a viscoelastic film of vanishing thickness within an infinite viscous
liquid, taking into account both in-plane and out-of-plane dynamics.
The in-plane response of this model, in the limit of a purely viscous,
incompressible film, coincides with that of the Saffman theory.

The third theoretical approach considers the membrane as an effective
2D Brinkman liquid \cite{Brinkman}, \ie a liquid with an additional
phenomenological term that makes it lose momentum over distances
larger than a certain given value, $\kappa^{-1}$
\cite{Evans&Sackmann,Suzuki&Izuyama,Seki&Komura,Komura&Seki,Seki&Komura2007}.  
The mobility of a disk of radius $a$, as calculated from this theory,
coincides with Eq.\ \ref{SD} in the limit $\kappa a\ll 1$.  However,
this approach is essentially different from the first two in that it
breaks the translational symmetry along the membrane surface.  Thus,
while the theories of Saffman and Levine \etal conserve total momentum
in 3D, allowing the surrounding liquid to impart momentum back to the
membrane at large distances, in the Brinkman-like theory the momentum,
once leaving the membrane, is lost.  As a result, the large-distance
response of this model is qualitatively different, decaying as $1/r^2$
(as required by mass conservation in 2D
\cite{IJC}) rather than $1/r$ (as resulting from momentum
conservation in 3D). This approach, therefore, is appropriate at
sufficiently short distances, or when translational symmetry is indeed
broken, as in membranes supported on a solid substrate
\cite{Evans&Sackmann,Stone1998}.

Since the lateral size of membrane proteins is typically at least one
order of magnitude larger than that of the lipids, they can be
considered as suspended in a continuous quasi-2D liquid, thus making
the membrane a quasi-2D suspension. In analogy with ordinary
suspensions, we expect that the average effect of many such mobile
proteins will lead to a modified effective response of the membrane.
For example, the presence of hard spheres in 3D suspensions leads to a
modified effective viscosity, which is given, up to linear order in
the volume fraction of spheres $\phi$, by \cite{Einstein}
\begin{equation}
  \mbox{3D suspension:}\ \ 
  \eta_{\rm f}^{\rm eff} = \etaf \left(1+\frac{5}{2}\phi\right).
\label{Einstein}
\end{equation}
The analogous result for a 2D suspension of hard disks is
\cite{Belzons}
\begin{equation}
  \mbox{2D suspension:}\ \ 
  \eta_{\rm m}^{\rm eff} = \etam \left(1+2\phi\right),
\label{Einstein2D}
\end{equation}
where here $\phi$ is the area fraction of disks. As described above,
membranes represent a more complicated intermediate between 2D and 
3D liquids, and we expect, therefore, a more subtle, distance-dependent
effect of inclusions on its response.

In the current work we extend the Saffman theory of membrane
hydrodynamics \cite{Saffman} to account for the correlated Brownian
motion of protein pairs and the effect of protein concentration on
that motion. The analysis is restricted to the limit of small protein
size, $\kappa a\ll 1$. Since the viscosity of a lipid bilayer is
typically $10^3$-fold that of water \cite{Dimova}, $\kappa^{-1}$ is
typically three orders of magnitude larger than the membrane
thickness, \ie of micron scale.  Hence, the limit $\kappa a\ll 1$
should hold for any membrane protein, as originally assumed by Saffman
and Delbr\"uck. As regards the inter-protein distance, two regimes are
addressed---intermediate distances, $a\ll r\ll\kappa^{-1}$, and the
asymptotically far region, $r\gg\kappa^{-1}$. We avoid the region
$r\sim a$, in which specific effects of protein shape and membrane
distortion are expected to be important. Corrections to leading order
in $a/r$ are nonetheless derived.

We begin with a presentation of several basic properties of Saffman's
hydrodynamic theory, which are useful for our calculations.  The
resulting coupling diffusion coefficients of an isolated protein pair
are subsequently described. We then proceed to derive the leading
corrections to the membrane response, as well as the coupling
diffusion coefficients, due to a low concentration of membrane
proteins.  Finally, we discuss the physical meaning of the results and
their practical limitations. While writing this article we have
learned of an independent study by Henle and Levine of the effective
viscosity of membranes with mobile inclusions \cite{Henle}. The
relation between their results and ours is addressed in the Discussion.

\section*{Theory}

\subsection*{Membrane hydrodynamics}

Consider a flat membrane, lying on the $xy$ plane. First, we address
the steady-state flow velocity of the membrane at position $\vecr$,
$\vecv(\vecr)$, in response to a point force, $\vecf$, applied at the
origin and directed along the membrane plane. In Fourier space,
$\tilde{\vecv}(\vecq)=\int d\vecr e^{-i\vecq\cdot\vecr}\vecv(\vecr)$,
Saffman's analysis yields \cite{Saffman}
\begin{equation}
  \tilde{v}_i(\vecq) = \tG_{ij}(\vecq)f_j,\ \ \ 
  \tG_{ij}(\vecq) = \frac{1}{\etam q(q+\kappa)} \left(\delta_{ij} - \frac{q_iq_j}{q^2}
  \right),
\label{OseenFourier}
\end{equation}
where $i,j=x,y$.  The tensor $\tenG$ is the membrane-analogue of the
Oseen tensor of Eq.\ \ref{Oseen3D}. Inverting to real space, we get
\begin{eqnarray}
  G_{ij}(\vecr) &=& \frac{1}{4\eta_m}\left\{
    \left[ H_0(\kappa r)-\frac{H_1(\kappa r)}{\kappa r}
    -\frac{1}{2}\left(Y_0(\kappa r)-Y_2(\kappa r)\right) +
    \frac{2}{\pi(\kappa r)^2} \right] \delta_{ij} \right.
 \nonumber\\
  && \left. - \left[ H_0(\kappa r)-\frac{2H_1(\kappa r)}{\kappa r} + Y_2(\kappa r)
  + \frac{4}{\pi(\kappa r)^2} \right] \frac{r_ir_j}{r^2} \right\},  
\label{Oseen}
\end{eqnarray}
where $Y_n$ and $H_n$ are, respectively, Bessel functions of the
second kind and Struve functions.  At short distances,
$r\ll\kappa^{-1}$, $\tenG$ reduces to
\begin{equation}
  r\ll\kappa^{-1}:\ \ G_{ij}(\vecr) \simeq 
  \frac{1}{4\pi\etam} \left\{ -\left[ \ln(\kappa r/2) + \gamma + 1/2 \right]
  \delta_{ij} + \frac{r_ir_j}{r^2} \right\}
  + O(\kappa r).
\label{smallOseen}
\end{equation}
This result coincides with the one for a 2D liquid, Eq.~\ref{Oseen2D},
with an appropriate definition of $\kappa'\sim\kappa$. At large
distances, $r\gg\kappa^{-1}$, $\tenG$ tends to
\begin{equation}
  r\gg\kappa^{-1}:\ \ G_{ij} \simeq \frac{1}{2\pi\etam}\frac{r_ir_j}{\kappa r^3}
 +O(\kappa r)^{-2}, 
\label{bigOseen}
\end{equation}
which shows the typical $1/r$ decay of 3D flows. Moreover, since
$\etam\kappa=2\etaf$, the large-distance response depends solely on
the outer liquid viscosity and is independent of membrane
viscosity. (In fact, up to a numerical prefactor, Eq.\ \ref{bigOseen}
could be readily derived by requiring that $\tenG$ decay as $1/r$ and
obey 2D incompressibility, $\pd_iG_{ij}=0$.)

Now suppose that, rather than being a point force, the force is
applied to the membrane by a disk-like particle of finite radius
$a$. To leading order in $a/r$, the membrane flow velocity is given by
$v_i(\vecr)\simeq G_{ij}(\vecr)f_j$.  We are interested in the
finite-size correction to this flow while still neglecting
terms of order $\kappa a$ (as we do throughout this work). The domain
of interest, therefore, is $a<r\ll\kappa^{-1}$. In this region the
membrane behaves as a 2D liquid, following Eq.~\ref{smallOseen}. Thus,
the calculation reduces to a 2D Stokes problem of finding the flow
away from a rigid disk, whose solution is
\begin{equation}
  a<r\ll\kappa^{-1}:\ \ \ v_i(\vecr) = G_{ij}^{(a)}(\vecr)f_j, \ \ \ 
  G_{ij}^{(a)}(\vecr) = G_{ij}(\vecr) + \frac{a^2}{8\pi\etam r^2}
  \left( \delta_{ij} - \frac{2r_ir_j}{r^2} \right),
\label{flow_a}
\end{equation}
where $\tenG$ is given by Eq.~\ref{smallOseen}.

Next we consider a membrane with a preexisting flow velocity
$\vecv(\vecr)$, and embed in it a circular disk of radius $a$ moving
with linear velocity $\vecU$ and angular velocity $\vecOmega$. We
would like to find the force $\vecF$, torque $\vecL$, and force dipole
(stresslet) $\tenS$, which the inclusion exerts on the fluid
membrane. For a sphere in an unbounded liquid the linear relations
between $(\vecF,\vecL,\tenS)$, on the one hand, and
$(\vecv,\vecU,\vecOmega)$, on the other, are given by the first and
second Fax\'en laws \cite{Happel}. In Appendix A we derive the
(approximate) membrane-analogues of these laws, which are as follows:
\begin{eqnarray}
  \vecF &\simeq& \frac{4\pi\etam} {\ln(\kappa a/2)+\gamma}
  \left( \vecv +\frac{1}{4}a^2\nabla^2\vecv - \vecU \right)
\label{Faxen1}\\
  \vecL &\simeq& 2\pi\etam a^2 \left[\left( 1 + \frac{1}{8}a^2\nabla^2 
  \right) \left(\nabla\times\vecv\right) - 2\vecOmega \right]
\label{Faxen2L} \\
  S_{ij} &\simeq& 2\pi\etam a^2 \left( 1 + \frac{1}{8}a^2\nabla^2 \right) 
  \left( \pd_i v_j + \pd_j v_i \right).
\label{Faxen2S}
\end{eqnarray}
Applying the first relation, Eq.~\ref{Faxen1}, to a disk moving in an
otherwise stationary membrane ($\vecv=0$), we recover the
Saffman-Delbr\"uck mobility (Eq.~\ref{SD}).  These relations are exact
for a 2D liquid (where, in addition, the term proportional to
$\nabla^2(\nabla\times\vecv)$ in Eq.\ \ref{Faxen2L} vanishes), whereas
for membranes their validity is restricted to sufficiently small
particles, $\kappa a\ll 1$.
(See Appendix A for details.)

\subsection*{Correlated diffusion}

Consider a pair of membrane proteins undergoing Brownian motion while
being separated by the 2D vector $\vecr$. We consider a time period
$t$, which is sufficiently short such that $\vecr$ can be assumed
constant, yet sufficiently long to yield Brownian displacements linear
in $t$. We further assume that $r$ is much larger than the protein
sizes (radii), $a_1$ and $a_2$.  The displacements of the two proteins
during time $t$ obey the following relations:
\begin{equation}
  \langle \Delta r_i^\alpha \Delta r_j^\beta \rangle
  = 2 D_{ij}^{\alpha\beta}(\vecr) t,
\label{diffusion}
\end{equation}
where $\Delta r_i^\alpha$ is the displacement of particle $\alpha$
($\alpha=1,2$) along the axis $i$ ($i=x,y$). The diffusion tensor
$D_{ij}^{\alpha\beta}$ characterizes both the self-diffusion of the
particles ($\alpha=\beta$) and the coupling between them
($\alpha\neq\beta$). We define the $x$ axis, without loss of
generality, along the line connecting the pair, \ie $\vecr=r\xhat$.
This choice leads, by symmetry, to $D_{xy}^{12}=0$. The coupled
diffusion is then fully characterized by two coefficients: a
longitudinal coupling diffusion coefficient,
$\Dl(r)=D_{xx}^{12}(r\xhat)$, and a transverse one,
$\Dt(r)=D_{yy}^{12}(r\xhat)$. The former is associated with the
coupled Brownian motion of the pair along their connecting line, and
the latter with the coupled motion perpendicular to that line.

For the overdamped dynamics considered here the diffusion tensor in
Eq.~\ref{diffusion} is simply related to a pair-mobility tensor via
the Einstein relation, $D_{ij}^{\alpha\beta}=\kT
B_{ij}^{\alpha\beta}$.  The mobility tensor $B_{ij}^{12}(\vecr)$ gives
the change in velocity $v_i$ of particle 2 located at $\vecr$ due to a
unit force in the $j$ direction applied to particle 1 at the origin.
In the limit $a/r\rightarrow 0$ ($a$ being the larger of $a_1,a_2$)
this, in turn, is just the membrane-analogue of the Oseen tensor,
Eq.~\ref{Oseen}. Hence, we readily identify,
\begin{eqnarray}
  &&r\gg(a^2\kappa^{-1})^{1/3}:\nonumber\\
  &&\Dl(r) \simeq \kT G_{xx}(r\xhat) = \frac{\kT}{4\etam\kappa r} 
  \left[ H_1(\kappa r) - Y_1(\kappa r)
      - \frac{2}{\pi\kappa r} \right] \nonumber\\
  &&\Dt(r) \simeq \kT G_{yy}(r\xhat) = \frac{\kT}{4\etam} 
  \left[ H_0(\kappa r)-\frac{H_1(\kappa r)}{\kappa r}
  - \frac{1}{2}\left(Y_0(\kappa r)-Y_2(\kappa r)\right)
  +\frac{2}{\pi(\kappa r)^2} \right]. 
\label{Dc}
\end{eqnarray}
(The domain of validity stated here will be clarified below, when we
address the effect of finite particle size.)  This result coincides
with the in-plane response functions derived by Levine and MacKintosh
in the limit of a purely viscous, incompressible membrane
\cite{Levine2002}.

At very large inter-particle distances, $r\gg\kappa^{-1}$, 
Eq.~\ref{Dc} reduces to
\begin{equation}
  r\gg\kappa^{-1}:\ \ 
  \Dl(r) \simeq \frac{\kT}{2\pi\etam\kappa r} = \frac{\kT}{4\pi\etaf r},\ \ 
  \Dt(r)\simeq \frac{\kT}{2\pi\etam(\kappa r)^2} = \frac{\kT\etam}{8\pi\etaf^2 r^2}.
\label{longDc}
\end{equation}
The longitudinal coupling between the two proteins decays
asymptotically as $1/r$ and is independent of membrane viscosity. (It
is identical, in fact, to the analogous coefficient in a 3D liquid.)
The transverse coupling decays only as $1/r^2$, and, curiously,
increases with membrane viscosity.
When the inter-particle distance is smaller than $\kappa^{-1}$ yet
still sufficiently large, Eq.~\ref{Dc} becomes
\begin{equation}
  (a^2\kappa^{-1})^{1/3}\ll r\ll\kappa^{-1}:\ \ \ 
  \Dlt(r) \simeq \frac{\kT}{4\pi\etam} \left[ -\ln(\kappa r/2) - \gamma \pm 1/2 
  +(1\mp 1/3)\kappa r \right],
\label{shortDc2}
\end{equation}
where the upper (lower) signs correspond to the longitudinal
(transverse) coefficient.
 
Let us now examine the leading effect of finite particle sizes, $a_1$
and $a_2$, and see at what inter-particle distance this effect becomes
significant. In this domain, clearly, $r\ll\kappa^{-1}$. First,
according to Eq.~\ref{flow_a} there is a correction of order
$a_1^2/r^2$ to the flow velocity caused by the forced particle 1.  In
addition, the first Fax\'en law, Eq.~\ref{Faxen1}, yields a correction
of order $a_2^2/r^2$ for the velocity of particle 2 as it is embedded
in that flow. Substituting $\vecv(\vecr)$ of Eq.~\ref{flow_a} in
Eq.~\ref{Faxen1} while setting $F=0$ (particle 2 being force-free), we
find the velocity $\vecU$ of particle 2 to leading order in
$a_1^2,a_2^2$. The relation between $\vecU$ and $\vecf$ defines a
corrected pair mobility tensor, resulting in the following coupling
diffusion coefficients:
\begin{equation}
  a\ll r\ll (a^2\kappa^{-1})^{1/3}:\ \ \ 
  \Dlt(r) \simeq \frac{\kT}{4\pi\eta_m} \left[ -\ln(\kappa r/2) 
  - \gamma \pm 1/2 \pm \frac{a_1^2+a_2^2}{2r^2}\right],
\label{shortDc1}
\end{equation}
where the plus (minus) sign corresponds to the longitudinal
(transverse) coefficient. The coefficients are symmetric under
particle exchange 1$\leftrightarrow$2, as they should be. Comparing
Eqs.~\ref{shortDc2} and \ref{shortDc1}, we see that the finite-size
effect sets in for $r\lesssim (a^2\kappa^{-1})^{1/3}$; hence the
domains of validity stated in Eqs.~\ref{Dc}--\ref{shortDc1}.

\begin{figure}[tbh]
\vspace{0.75cm} 
\centerline{\resizebox{0.45\textwidth}{!}{\includegraphics{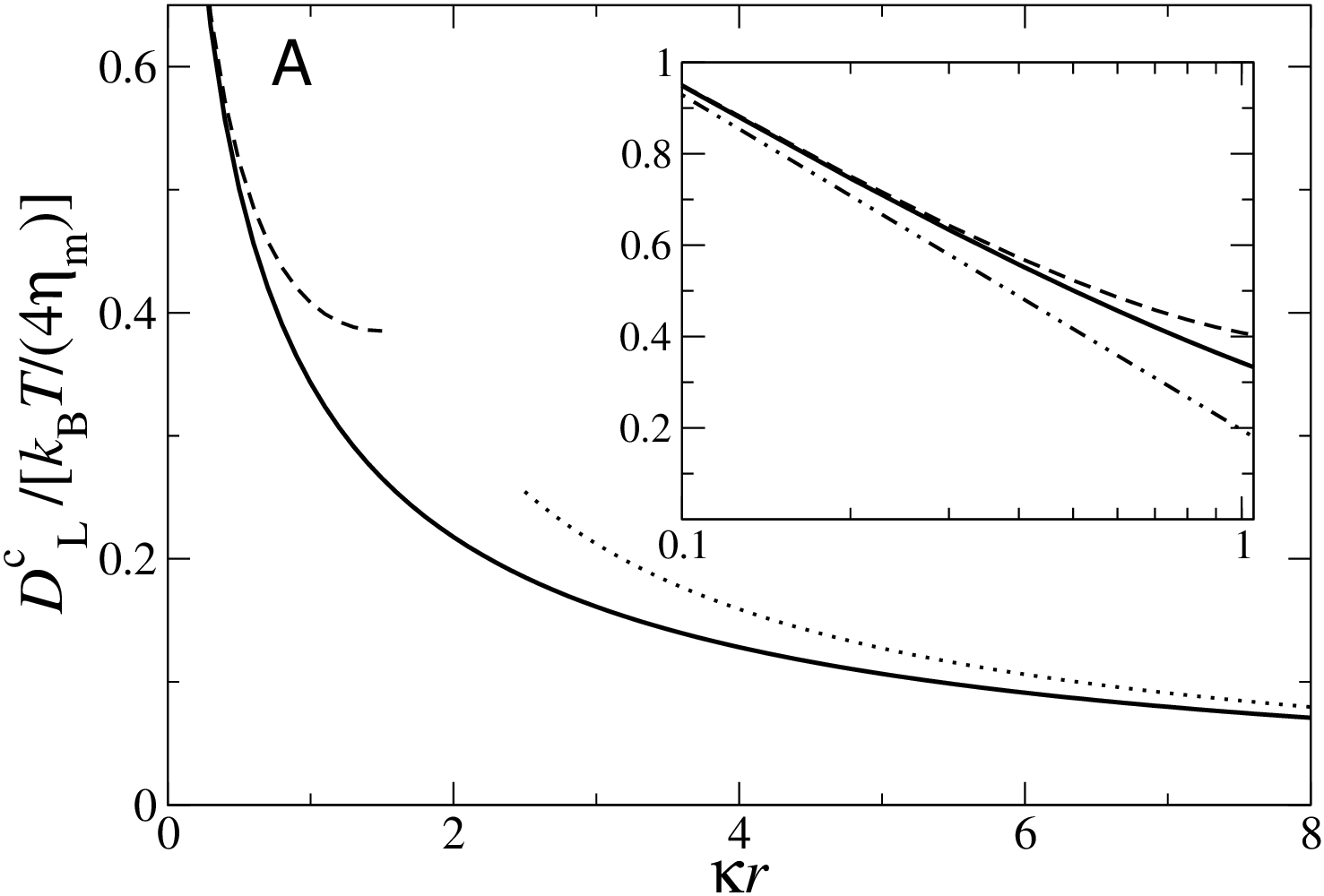}}
\hspace{0.1cm}
\resizebox{0.45\textwidth}{!}{\includegraphics{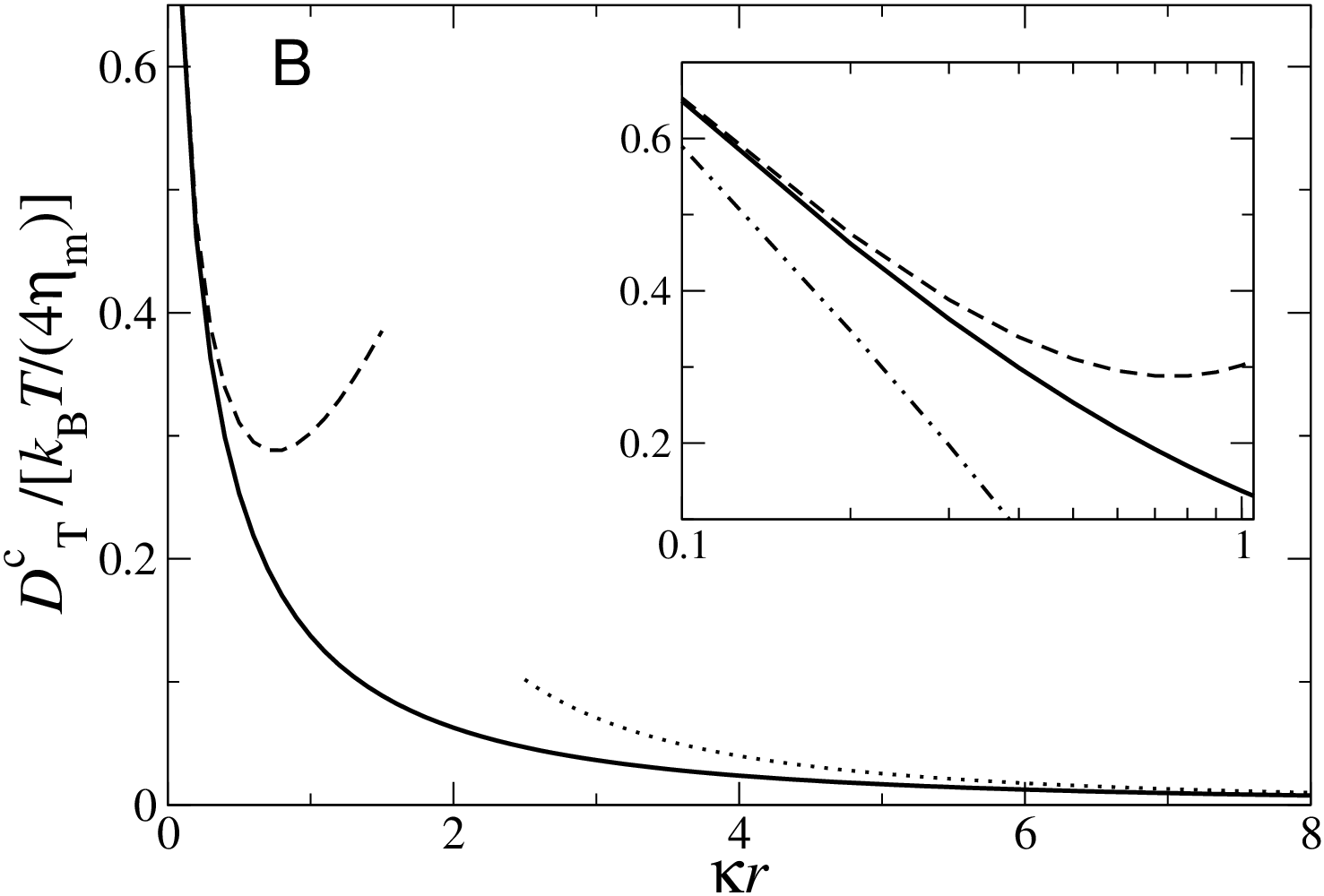}}}
\caption{\small Longitudinal (A) and transverse (B) coupling diffusion
coefficients as a function of inter-protein distance. The diffusion
coefficients are scaled by $\kT/(4\etam)$ and the distance by the
Saffman-Delbr\"uck length $\kappa^{-1}$. The results for
$r\gg(a^2\kappa^{-1})^{1/3}$ (Eq.\ \ref{Dc}, solid lines) are
presented along with their asymptotes for $r\gg\kappa^{-1}$ (Eq.\
\ref{longDc}, dotted lines) and $(a^2\kappa^{-1})^{1/3}\ll r\ll
\kappa^{-1}$ (Eq.\ \ref{shortDc2}, dashed lines). The insets focus on
the region of shorter distances, where corrections due to protein size
become significant (Eq.\ \ref{shortDc1}, dash-dotted lines, taking
$a_1=a_2=10^{-3}\kappa^{-1}$).}
\label{fig_Dc}
\end{figure}

Figure~\ref{fig_Dc} shows the coupling diffusion coefficients as a
function of inter-particle distance, along with their asymptotes.

\subsection*{Effective viscosity}

We would like now to calculate the change in membrane viscosity due to
the presence of many embedded proteins, to leading (linear) order in
protein concentration. The definition of effective viscosity is a
subtle point, to which we return in the Discussion. Here we define it
using the large-distance flow response of the membrane, \ie we extract
the effective viscosity from
$\tenG(r\rightarrow\infty)\rightarrow\tenG^{\rm eff}$, as it is
modified by protein concentration. A similar procedure for a 3D
suspension of hard spheres \cite{IJC} correctly reproduces the known
effective viscosity of that system, Eq.~\ref{Einstein}.

We begin again by applying a point force $\vecf$ at the origin. The
membrane flow velocity at position $\vecr$ is then given by
$v^{(0)}_i(\vecr)=G_{ij}(\vecr)f_j$, where $\tenG$ is given by
Eq.~\ref{Oseen}.  Let a disk-like protein of radius $a$ be positioned
at $\vecr'$. Due to its finite size it will perturb the membrane flow.
The particle is force- and torque-free, and, hence, the leading moment
of that perturbation is a force dipole, $\tenS(\vecr')$.  From our
second Fax\'en-like law, Eq.~\ref{Faxen2S}, we have
\begin{equation}
  S_{ij}(\vecr') = 2\pi\etam a^2 [\pd_i G_{jk}(\vecr') + \pd_j G_{ik}(\vecr')] f_k,
\label{Sij}
\end{equation}
where we have neglected terms of order $a^4$.  This force dipole,
located at $\vecr'$, changes the flow velocity at position $\vecr$ by
$\delta v_i(\vecr)=S_{kj}(\vecr')\pd_k G_{ij}(\vecr-\vecr')$.

Now suppose that many such mobile proteins are present in the
membrane, occupying a fraction $\phi$ of its area.  The theory being
linear, we can superimpose their individual perturbations and average
over all possible positions $\vecr'$. This yields an average
correction to the flow velocity,
\begin{equation}
  \langle\delta v_i(\vecr)\rangle = \int d\vecr' p(\vecr') S_{kj}(\vecr')\pd_k G_{ij}(\vecr-\vecr'),
\label{MeanCorrection}
\end{equation} 
where $p(\vecr')$ is the probability density of finding a particle at
$\vecr'$.  To leading order in $\phi$ we may assume a uniform
probability density, $p(\vecr')=\phi/(\pi a^2)$.  The convolution in
Eq.~\ref{MeanCorrection} is then conveniently handled in Fourier
space, $\langle\delta\tilde{v}_i(\vecq)\rangle=[\phi/(\pi
a^2)]\tilde{S}_{kj}(\vecq)iq_k\tG_{ij}(\vecq)$.  Substituting
Eq.~\ref{OseenFourier} and the transform of Eq.~\ref{Sij}, we find
\begin{equation}
  \tilde{v}_i(\vecq) = \tilde{v}^{(0)}_i + \langle\delta\tilde{v}_i\rangle
  = \tG^{\rm eff}_{ij}(\vecq) f_j,\ \ \ 
  \tilde{\tenG}^{\rm eff} = \left(1 - \frac{2\phi q}{q+\kappa}\right)\tilde{\tenG}.
\label{GeffFourier}
\end{equation}
In the limit $\kappa\rightarrow 0$ Eq.\ \ref{GeffFourier} reduces to
the effective response of a 2D suspension of hard disks,
Eq.~\ref{Einstein2D} \cite{Belzons}. For finite $\kappa$, because of
the $q$-dependent prefactor in Eq.~\ref{GeffFourier}, it does not seem
at first as if the mobile particles could lead to such a
straightforward renormalization of the membrane response, as they do
in 2D and 3D suspensions \cite{IJC}.  Only in the limit
$q\rightarrow\infty$ do we simply get $\tenG^{\rm
eff}\rightarrow(1-2\phi)\tenG$ as in the 2D case. In the opposite limit,
$q\rightarrow 0$, we obtain $\tenG^{\rm eff}\rightarrow\tenG$, \ie the
membrane response becomes unaffected by the presence of proteins.  A
closer inspection reveals, however, that Eq.~\ref{GeffFourier} could
be also obtained from $\tilde{\tenG}$ of Eq.~\ref{OseenFourier} by the
following simple substitution (up to linear order in $\phi$):
\begin{equation}
  \tG^{\rm eff} = \tG|_{\etam\rightarrow\etam(1+2\phi),\kappa\rightarrow\kappa(1-2\phi)}.
\end{equation}
Furthermore, since $\kappa=2\etaf/\etam$, the two substitutions (to
linear order in $\phi$) are one and the same. Thus, as in a 2D
suspension of hard disks \cite{Belzons}, one can write the effective viscosity of the
membrane as
\begin{equation}
  \etaeff = \etam(1+2\phi),
\label{etaeff}
\end{equation}
provided that this modification is applied to the Saffman-Delbr\"uck
length as well.

One should not be confused by the similarity of Eqs.\ \ref{Einstein2D}
and \ref{etaeff}; the effective hydrodynamic response of a
protein-laden membrane is not at all similar to that of a 2D
suspension.  At short distances (yet still much larger than the
protein size), substituting $\etam\rightarrow\etam(1+2\phi)$ and
$\kappa\rightarrow\kappa(1-2\phi)$ in Eq.~\ref{smallOseen} leads to
\begin{equation}
  r\ll\kappa^{-1}:\ 
  G^{\rm eff}_{ij}(\vecr) \simeq (1-2\phi)G_{ij}(\vecr) 
  + \frac{\phi}{2\pi\eta_m} \delta_{ij}.
\end{equation}
Thus, even in this region of 2D-like behavior, there is an extra term
in the membrane response, arising from the modification of
$\kappa$. The effect becomes much more dramatic in the large-distance
limit, where we have from Eq.~\ref{bigOseen}
\begin{equation}
  r\gg\kappa^{-1}:\
  G_{ij}^{\rm eff}(\vecr)\simeq G_{ij}(\vecr),
\end{equation}
without any effect of the embedded proteins. As in Eq.~\ref{bigOseen}
the underlying physics is that over large distances stresses are
transmitted through the surrounding liquid.  Hence, the response
becomes indifferent to the properties of the membrane, be it with or
without proteins.

\subsection*{Concentration corrections to pair diffusion}

\begin{figure}[tbh]
\vspace{0.75cm} 
\centerline{\resizebox{0.48\textwidth}{!}{\includegraphics{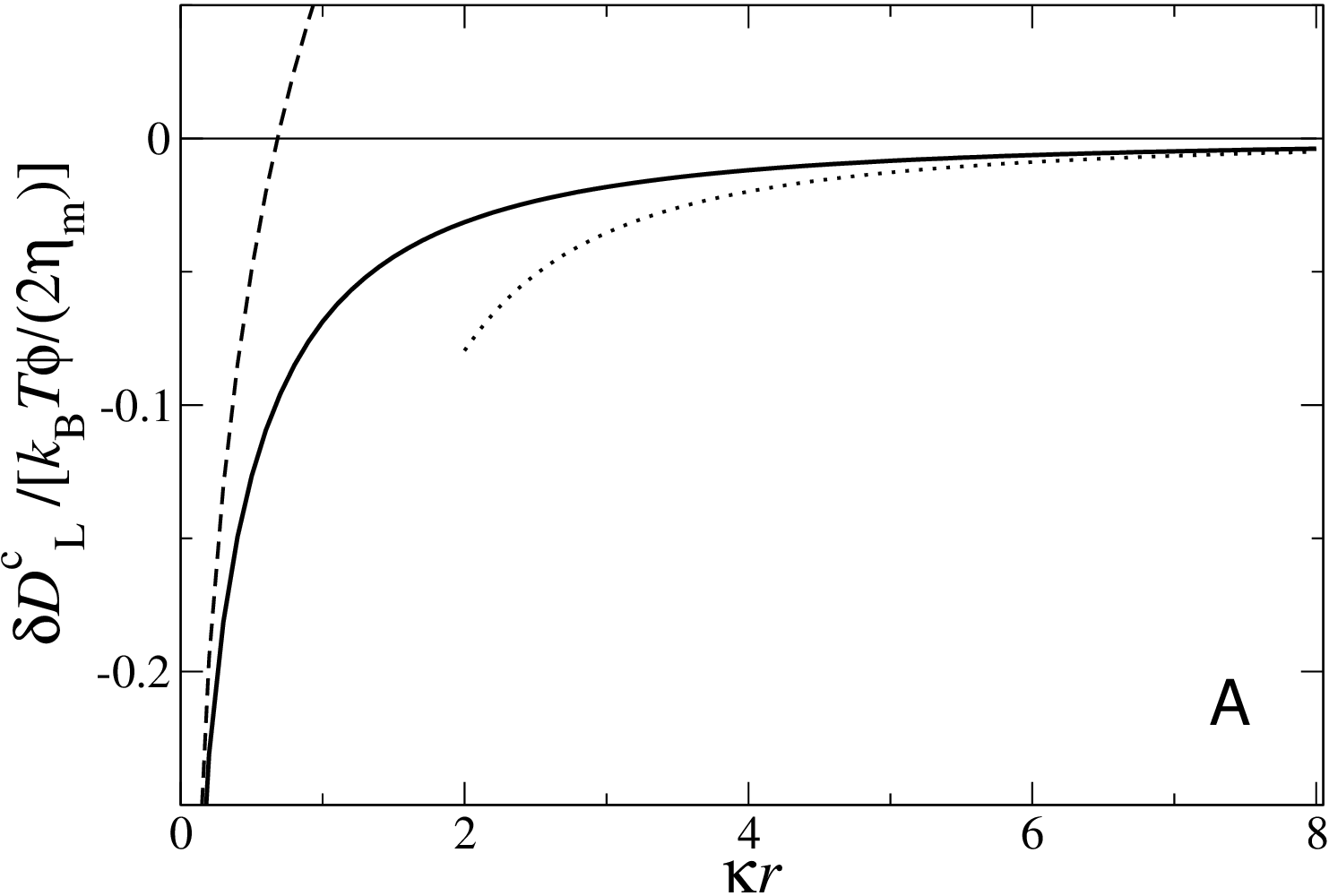}}
\hspace{0.2cm}
\resizebox{0.48\textwidth}{!}{\includegraphics{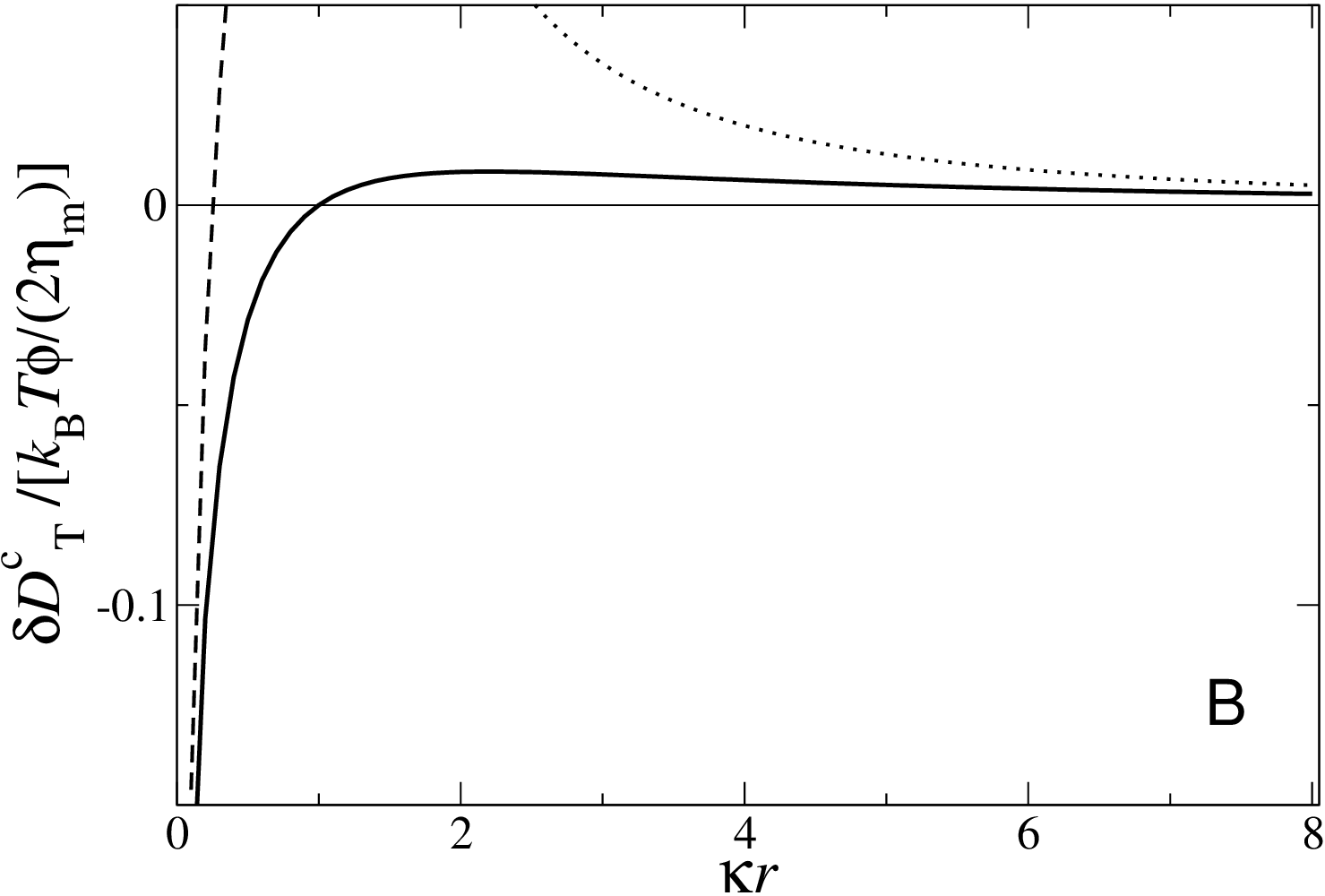}}}
\caption[]{\small Concentration corrections to the longitudinal 
 (A) and transverse (B) coupling diffusion coefficients as a function
 of inter-protein distance. The corrections to the diffusion
 coefficients are scaled by $\kT\phi/(2\etam)$ and the distance by the
 Saffman-Delbr\"uck length $\kappa^{-1}$. The results for
 $r\gg(a^2\kappa^{-1})^{1/3}$ (Eq.\ \ref{deltaDc}, solid lines) are
 presented along with their asymptotes for $r\gg\kappa^{-1}$ (Eq.\
 \ref{bigdeltaDc}, dotted lines) and $(a^2\kappa^{-1})^{1/3}\ll r\ll
 \kappa^{-1}$ (Eq.\ \ref{smalldeltaDc}, dashed lines).}
\label{fig_delta}
\end{figure}

The results of the preceding section can be readily used to obtain the
corrections to the coupling diffusion coefficients due to the presence
of many mobile, disk-like proteins, occupying an area fraction $\phi$.
All we need to do is substitute in Eq.~\ref{Dc}
$\etam\rightarrow\etam(1+2\phi)$, $\kappa\rightarrow\kappa(1-2\phi)$,
and expand to linear order in $\phi$. This results in
\begin{eqnarray} 
  &&r\gg(a^2\kappa^{-1})^{1/3}:\nonumber\\
  &&\delta\Dl(r) = -\phi \frac{\kT}{2\etam} \left[ 
  H_0(\kappa r)-\frac{H_1(\kappa r)}{\kappa r} + \frac{1}{2}(Y_2(\kappa r)-Y_0(\kappa r))
  + \frac{2}{\pi(\kappa r)^2}\right] \nonumber\\
  &&\delta\Dt(r) = -\phi\frac{\kT}{2\etam} \frac{(\kappa r)^2-1}{\kappa r}
  \left[H_{-1}(\kappa r) + Y_1(\kappa r) +
  \frac{2}{\pi\kappa r(\kappa r+1)} \right].
\label{deltaDc}
\end{eqnarray}
Equation \ref{deltaDc} gives the concentration corrections to the bare
coupling diffusion coefficients given in Eq.~\ref{Dc}. Their spatial
dependencies are depicted in Fig.~\ref{fig_delta}. The correction to
the longitudinal coupling is always negative, whereas the correction
to the transverse one becomes positive for $r>\kappa^{-1}$. This is
because at such large distances the bare transverse coefficient $\Dt$
(Eq.~\ref{longDc}) increases, rather than decreases, with $\etam$.

At large distances we get the following corrections to Eq.~\ref{longDc}:
\begin{equation} 
  r\gg\kappa^{-1}:\ \ \ 
  \delta\Dlt(r) \simeq \mp \phi \frac{\kT}{\pi\etam (\kappa r)^2}. 
\label{bigdeltaDc}
\end{equation}  
Recall that the bare longitudinal coefficient decays at large
distances as $1/r$, whereas the transverse one decays as $1/r^2$
(Eq.~\ref{longDc}). Hence, according to Eq.~\ref{bigdeltaDc}, the
asymptotic behavior of $\Dl$ is unaffected by the presence of the
proteins, $\Dl\rightarrow\Dl$. This is because at such distances the
bare $\Dl$ is independent of membrane viscosity (Eq.~\ref{longDc}).
By contrast, the transverse coefficient is affected (increased) by
$\phi$ at large distances, $\Dt\rightarrow(1+2\phi)\Dt$.

At distances much shorter than $\kappa^{-1}$ Eq.~\ref{deltaDc} becomes
\begin{eqnarray}
  (a^2\kappa^{-1})^{1/3}\ll r\ll\kappa^{-1}:\ \ \  
  \delta\Dlt(r) \simeq \phi \frac{\kT}{2\pi\etam} 
  \left[ \ln(\kappa r/2) + \gamma + 1 \mp 1/2 \right].
\label{smalldeltaDc}
\end{eqnarray}
Comparing with Eq.~\ref{shortDc2}, we find that the leading term
[$\sim\ln(\kappa r)$] is renormalized by $\phi$ as in a 2D liquid,
$\Dlt\rightarrow(1-2\phi)\Dlt$. Yet, the next-order term [$O(1)$] does
not follow the same law. This is again because there is another
dependence on $\etam$ in the length $\kappa^{-1}$.

\section*{Discussion}

The theory presented here has yielded several predictions concerning
the correlated Brownian motion of pairs of membrane proteins, which
can be directly checked in two-particle tracking experiments using
Eq.~\ref{diffusion}. Equation \ref{Dc} gives the longitudinal and
transverse coupling diffusion coefficients as a function of
inter-protein distance. An equivalent result, in the form of in-plane
response functions, has been previously reported \cite{Levine2002}. It
is valid at sufficiently large distances and insensitive to the size
and shape of the proteins. For smaller distances, yet still larger
than the protein size $a$, we have derived expressions for the
coupling diffusion coefficients to leading order in $a/r$ and assuming
a disk-like shape of the embedded proteins (Eq.~\ref{shortDc1}). Since
the finite-size correction is only quadratic in $a/r$, it becomes
significant only for distances $r\lesssim(a^2\kappa^{-1})^{1/3}$.  For
$a\sim 1$ nm and $\kappa^{-1}\sim 10^3$ nm this crossover length is
only $\sim 10$ nm.  Thus, we expect the finite size of proteins to
affect their hydrodynamic coupling only at nanometer-scale
distances. We have provided the corrections to the two coupling
coefficients due to the presence of other membrane proteins, to
leading order in their area fraction $\phi$ (Eq.~\ref{deltaDc}).

Several particular predictions are worth emphasizing.  First, the
longitudinal diffusion coefficient decays with distance more slowly
than the transverse one. Asymptotically, $\Dl$ decays as $1/r$,
whereas $\Dt$ decays as $1/r^2$ (Eq.~\ref{longDc}). Second, at such
large distances $\Dl$ becomes independent of membrane viscosity and
is, in fact, identical to the longitudinal coefficient in an
unbounded liquid.  This is because the coupling in this regime is
mediated by flows in the surrounding liquid. Thus, the large-distance
longitudinal coupling should be the same for different membranes in
the same solvent and can be tuned by changing the solvent
viscosity. The dominance of stresses in the outer liquid holds for the
transverse coupling as well, yet this coupling arises from an
effective force dipole, which is proportional to
$\kappa^{-1}=\etam/(2\etaf)$ and, therefore, remains
membrane-dependent.  We note that these asymptotes hold for
$r\gg\kappa^{-1}\sim 0.1$--$1$ $\mu$m and, hence, may be hard to
observe in practice. Nonetheless, the difference in the spatial decays
should be seen already at much shorter distances. (See
Figs.~\ref{fig_Dc}A and \ref{fig_Dc}B.) Third, the longitudinal
coefficient at large distances is predicted to be independent of
protein concentration (Eq.~\ref{bigdeltaDc}). This is merely another
consequence of the membrane-independence of this coefficient.
By contrast, the large-distance transverse coefficient not only
depends on protein concentration but increases with $\phi$
(Eq.~\ref{bigdeltaDc} and Fig.~\ref{fig_delta}B). This unusual
result---hydrodynamic interaction enhanced by particle
concentration---stems from the aforementioned effective force dipole,
which increases with $\kappa^{-1}\sim\etam$.

The concept of effective viscosity 
may have different, not necessarily equivalent, definitions
\cite{Happel}. The case of a membrane with mobile inclusions seems to
clearly demonstrate this difficulty. If one were to measure $\etaeff$
from the large-distance longitudinal coupling coefficient as a
function of $\phi$, one would find no concentration effect,
$\etaeff=\etam$. If, however, $\etaeff$ were extracted from the
large-distance transverse coefficient, a concentration dependence
identical to that in a 2D liquid would follow,
$\etaeff=(1+2\phi)\etam$.  (Compare Eqs.~\ref{longDc} and
\ref{bigdeltaDc}.) The calculation by Henle and Levine \cite{Henle} is
based on the same (Saffman) theory, yet follows Einstein's original
definition of the effective viscosity as the coefficient relating
average stress (or dissipation rate) with strain rate under a given
global shear flow \cite{Einstein}. They find yet another concentration
dependence in the limit $\kappa a\rightarrow 0$,
$\etaeff=\etam(1+3\phi)$.  In the cases of 3D and 2D suspensions, as
shown, respectively, in Ref.\ \cite{IJC} and the current work, the
aforementioned three definitions of $\etaeff$ give identical
results. The different behavior of membranes lies in the fact that
they do not conserve momentum, \ie in the appearance of the length
scale $\kappa^{-1}$ and its dependence on $\etam$.

Thus, the effect of inclusions on membrane viscosity depends on the
definition of that transport coefficient and the experiment under
consideration.  The definition used here relates to how the response
of the membrane to local perturbations changes with $\phi$.  It is
relevant, therefore, to microrheological and particle-tracking
experiments. Einstein's definition, as used by Henle and Levine
\cite{Henle}, should be appropriate for larger-scale rheological
measurements. We have found that the effective response in the
presence of proteins is given, as in a 2D suspension, by substituting
$\etam\rightarrow\etaeff=\etam(1+2\phi)$.  Yet, unlike 2D suspensions,
this substitution should be made also in the Saffman-Delbr\"uck
length, $\kappa^{-1}=\etam/(2\etaf)\rightarrow\etaeff/(2\etaf)$.  The
extra dependence of $\kappa$ on $\phi$ leads to a qualitatively
different effective response. A clear demonstration is given by the
concentration correction to the transverse coupling, $\delta\Dt$,
where the interplay between the $\phi$-dependencies of $\etam$ and
$\kappa$ leads to a sign reversal of that term (Fig.\
\ref{fig_delta}B). Another consequence is the aforementioned
independence of the large-distance longitudinal coupling, $\Dl$, on
protein concentration.

We note that a similar absence of a concentration effect on the
large-distance response has been observed in another quasi-2D
system---a suspension confined between two plates
\cite{Cui2004,Cui2005}. The physical origins of the two phenomena,
however, are slightly different. In the confined suspension momentum
is lost to the solid boundaries, and the far response arises solely
from liquid mass displacement, which is not affected by the presence
of particles.  In membranes the far response does arise from momentum
diffusion, yet these dynamics take place in the outer liquid and,
therefore, are insensitive to the presence of membrane inclusions.

The current analysis has involved several rather severe
approximations.  It should be regarded, therefore, as a first step
toward understanding the correlated dynamics of membrane proteins or,
alternatively, as a possible means to isolate simple hydrodynamic
effects from other, more specific ones.  First, we have focused on the
hydrodynamic coupling between proteins, neglecting any direct
interaction \cite{Abney}. Such interactions may arise from actual (\eg
electrostatic) potentials or be induced by the perturbations that
inclusions introduce in the membrane
\cite{Dan1993,Goulian1996,Naji2007}. Second, as in previous theories,
we have considered a homogeneous membrane, whereas actual biomembranes
are believed to contain various heterogeneities and domains
\cite{Kusumi}. A homogeneous hydrodynamic description may still be
applicable inside such a sub-micron domain. Third, the calculations
have been made in the limit of very small particle size, $\kappa a\ll
1$.  As the typical values of $\kappa^{-1}$ are micron-scale, this
should be a good approximation for practically all membrane proteins.
The Saffman theory can be extended to large values of $\kappa a$ as
well, yet the calculations become significantly more complicated
\cite{Hughes1980,Henle}.  Fourth, we have treated the membrane as a
perfectly flat surface, whereas in practice it is curved and
fluctuating. Curvature and bending fluctuations, apart from their
aforementioned ability to induce interactions between embedded
proteins, may also affect their 2D-projected diffusion as observed in
experiments \cite{Seifert2007,Seifert2008}.  

Finally, we have studied the effect of protein concentration to linear
order only. As in the much simpler case of a 3D suspension of hard
spheres, extension to higher orders in $\phi$ should be difficult,
involving static and dynamic correlations between particles.
Nevertheless, some of our results clearly emanate from more
fundamental considerations and, therefore, should hold for higher
values of $\phi$ as well.  For example, the $\phi$-independence of
$\Dl(r\gg\kappa^{-1})$ arises from stresses being transmitted through
the outer liquid; thus, we conjecture that it is valid to all orders
in $\phi$. It is also plausible that, in the limit $\kappa a\ll 1$,
expressions of higher order in $\phi$ could be obtained by merely
substituting for $\etam$ (both directly and in $\kappa$) the effective
viscosity, $\etaeff(\phi)$, as calculated for a 2D suspension.

\section*{Appendix A: Fax\'en laws for a membrane}

In this appendix we calculate the approximate membrane-analogue of the
3D Fax\'en laws. These laws relate the linear velocity $\vecU$ and
angular velocity $\vecOmega$ of a rigid particle, and the flow field
$\vecv(\vecr)$ in which it is embedded, with the moments of force
distribution that it exerts on the embedding fluid.

Let the center of a disk of radius $a$ be located at the origin, and
let $\vecf(\vecr')$ be the distribution of forces that it exerts on
the system.  This force distribution changes the membrane velocity at
position $\vecr$ by $\int d\vecr' G_{ij}(\vecr-\vecr')f_j(\vecr')$,
where $\tenG$ is the membrane-analogue of the Oseen tensor, Eq.\
\ref{Oseen}.  Assuming no slip at the particle perimeter, we have
\begin{equation}
  r=a:\ \ U_i + (\vecOmega\times\vecr)_i = v_i(\vecr) 
  + \int d\vecr' G_{ij}(\vecr-\vecr')f_j(\vecr').
\label{eqSuperposition}
\end{equation}
We have intentionally left the domain $\vecr'$ of the force
distribution unspecified, since the calculation is insensitive to it;
the vector $\vecr$, however, must be on the particle perimeter, where
the no-slip boundary condition is imposed.

Assuming that $\vecv(\vecr)$ changes very moderately on the scale of
$a$, we expand $\vecv(\vecr)\simeq\vecv({\bf 0}) +
a\rhat_i\pd_i\vecv({\bf 0}) +
\frac{1}{2}a^2\rhat_i\rhat_j\pd_i\pd_j\vecv({\bf 0})$. Since both $r$
and $r'$ in Eq.~\ref{eqSuperposition} are of order $a$, and we have
been assuming throughout this work $\kappa a\ll 1$, we can substitute
for $\tenG$ its short-distance asymptote, Eq.~\ref{smallOseen}. Within
these approximations, integrating Eq.~\ref{eqSuperposition} over
$\vecr$ yields the analogue of the first Fax\'en law,
\begin{equation}
  \vecF = \frac{4\pi\etam} {\ln(\kappa a/2)+\gamma + O(\kappa a)^2}
  \left[ \vecv({\bf 0}) +\frac{1}{4}a^2\nabla^2\vecv({\bf 0}) - \vecU
  + O(a^4\nabla^4\vecv) \right],
\label{Faxen1app}
\end{equation}
where $\vecF=\int d\vecr' \vecf(\vecr')$.

Next we multiply both sides of Eq.~\ref{eqSuperposition} by $\vecr$
and integrate over $\vecr$. Separating the resulting tensors into
symmetric and antisymmetric contributions, we obtain the analogue of
the second Fax\'en law for the torque and force dipole (stresslet),
\begin{eqnarray}
  &&\vecL = 2\pi \etam a^2 [1 + O(\kappa a)^2] \left\{ \left[ 
  1 + \frac{1}{8}a^2\nabla^2 
  + O(a^4\nabla^4) \right] \left[\nabla\times\vecv({\bf 0})\right] - 
  2\vecOmega \right\}
\label{Faxen2Lapp} \\
  &&S_{ij} = 2\pi\etam a^2[1+O(\kappa a)^2] 
  \left[ 1 + \frac{1}{8}a^2\nabla^2 + O(a^4\nabla^4)\right] 
  \left[ \pd_i v_j({\bf 0}) + \pd_j v_i({\bf 0}) \right],
\label{Faxen2Sapp}
\end{eqnarray}
where $\vecL=\int d\vecr' [\vecr'\times\vecf(\vecr')]$ and
$S_{ij}=(1/2)\int d\vecr' [r'_i f_j(\vecr')+r'_j f_i(\vecr')]$.

In regular Stokes flows $\kappa=0$, $\nabla^4\vecv=0$,
$\nabla^2(\nabla\times\vecv)=0$, and the Fax\'en laws become
exact. Thus, Eqs.~\ref{Faxen1app}--\ref{Faxen2Sapp} are exact for 2D
liquids, where, additionally, the term proportional to
$\nabla^2(\nabla\times\vecv)$ in Eq.\ \ref{Faxen2Lapp} vanishes.  For
membranes, however, their validity is restricted to sufficiently small
particles, $\kappa a\ll 1$. In addition, the terms of order $a^2$ are
valid provided that the considered flow is not too uniform, $|\nabla^2
v/v| \gg \kappa^2$.

\section*{Acknowledgments}

We thank S.\ Komura for helpful discussions. This work has been
supported by the Israel Science Foundation (Grant no.\ 588/06).

 \end{document}